\documentclass[floatfix,twocolumn,showpacs,preprintnumbers,amsmath,amssymb,superscriptaddress]{revtex4}

\usepackage{hyperref}
\usepackage{graphicx}

\newcommand{\ket}[1]{\ensuremath{| #1\rangle}}

\renewcommand{\tensor}{\otimes}

\begin{document}

\title{
    Single Spin Measurement Using Cellular Automata Techniques
}

\author{Carlos A. P\'{e}rez-Delgado}
\affiliation{%
Institute for Quantum Computing,
University of Waterloo,
Waterloo, ON N2L 3G1, Canada
}%

\author{Michele Mosca}%
\affiliation{%
Institute for Quantum Computing,
University of Waterloo,
Waterloo, ON N2L 3G1, Canada
}%
 \affiliation{
 Perimeter Institute for Theoretical Physics,
 Waterloo, ON N2J 2W9, Canada}

\author{Paola Cappellaro}
\affiliation{
Department of Nuclear Engineering,
Massachusetts Institute of Technology,
Cambridge, MA 02139, USA
}%

\author{David G. Cory}
\affiliation{
Department of Nuclear Engineering,
Massachusetts Institute of Technology,
Cambridge, MA 02139, USA
}%

\date{\today}


\begin{abstract}
We propose an approach for single spin measurement.  Our method uses techniques from the theory of quantum cellular automata to correlate a large amount of ancillary spins to the one to be measured. It has the distinct advantage of being efficient, and to a certain extent fault-tolerant. Under ideal conditions, it requires the application of only $O(\sqrt[3]{N})$ steps (each requiring a constant number of rf pulses)  to create a system of $N$ correlated spins. It is also fairly robust against pulse  errors, imperfect initial polarization of the ancilla  spin system, and does not rely on entanglement. We study the scalability of our scheme through numerical simulation.
\end{abstract}

\maketitle

One of the most interesting challenges in physics today is that of measuring the state of a single (nuclear) spin. Being able to do this would bring us closer to spin based quantum computers \cite{kane1998}, and have a myriad of applications, ranging from spintronics, to protein analysis. Unfortunately, performing such measurement is not an easy task. Several methods have been proposed \cite{wrachtrup1997}, and single-spin \emph{detection} has been done \cite{rugar2004}, and the measurement in some specific cases has also been achieved \cite{wrachtrup2002}.

 Cappellaro \emph{et. al.}  \cite{paola} propose  using a system of $N$ independent spins as a measurement system. This system is coupled to the spin being measured, and using entangling operations, creates a large correlated  state, one whose signal can be measured with current NMR technology. The methods presented there require $O(N)$ pulse sequences in order to achieve $N$ quanta of polarization. We expand on their ideas to create a scheme which uses $O(\sqrt[3]{N})$ pulses, and is fairly robust against noise and errors, and does not rely on entanglement. The improved running time is the most important advantage, since the whole procedure must finish before decoherence destroys the information being measured.
  
\begin{figure}
\includegraphics[scale=1]{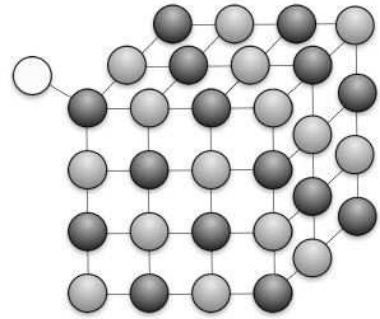}
\caption{\label{cube} \emph{Cube lattice:} A crystal with two types of nuclei $A$ and $B$, one represented as light gray spheres, the other dark gray. Each species $A$ is neighbored by only $B$ type nuclei and vice-versa. The lines connecting the spheres represent the nearest neighbor couplings. The white sphere represents the spin  we wish to measure. This spin is coupled to the dark gray nucleus in the top-left vertex. }
\end{figure}

The method presented here is inspired by quantum cellular automata \cite{watrous,carlos1} and pulse driven quantum computers \cite{lloyd93,benjamin1}. It uses a cubic lattice crystal \footnote{A cube lattice is not required. We chose a cube here as it allows for the simplest possible exposition. Later we show that using a rhombohedral lattice is more appropriate when dealing with solid state NMR.} with two nucleus types, which we call $A$ and $B$. We assume, for the time being, only nearest neighbor interaction. Each species $A$ nucleus is connected only to $B$ nuclei, and \emph{vice versa} in a checkerboard fashion. See fig. \ref{cube}

We will refer to upward and downward $z$ polarizations as $\ket{+1}$ and $\ket{-1}$. 
We assume, for the time being,  that the crystal is initialized to a completely polarized state with all nuclear spins in a downwards $z$ polarization state, $\ket{ -1}$. The method then consists of bringing one corner of the crystal into close proximity to the spin we wish to measure, so as to couple the two spins. Once coupled we can use NMR rf pulses to correlate them, or swap the states. Suppose the spin we wish to measure is initially in the state  $\ket{\psi} = \alpha \ket{-1} + \beta \ket{+1}$. After the swap the top-left vertex nuclear spin will be in the state $\ket{\psi}$.

What we present now is an efficient method to create a very large correlated state within the crystal. Under ideal conditions, that is complete polarization, perfect pulses, and no decoherence, the method  creates the state  $\alpha \ket{-1}^{\tensor N} + \beta \ket{+1}^{\tensor N} $ using only $O(\sqrt[3]{N})$ steps (each requiring a constant number of rf pulses). When $N \sim 10^6$ the resulting state gives a strong enough magnetic signal to be measured. Achieving this polarization, in the ideal case, requires applying  about 200 steps of our algorithm. 

\begin{figure}
\includegraphics[scale=1]{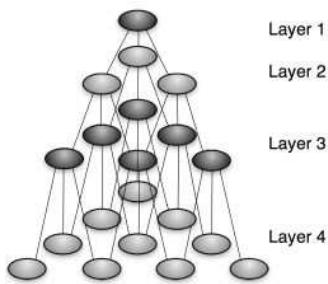}
\caption{\label{pyramid} \emph{Pyramid lattice:} A different view on the same structure as fig. \ref{cube}. The top gray sphere is layer 1 and corresponds to the top-left vertex spin in fig. \ref{cube}. The three light gray spheres directly below it are layer two, and so on. The lines connecting the spheres represent the nearest neighbor couplings, and are exactly the same as in fig. \ref{cube}.}
\end{figure}

In order to more easily illustrate our algorithm it is best to visualize the cube lattice in the following way. We envision slicing the cube into layers, such that the first layer is the corner nuclear spin that contains the state to be measured. Layer two contains all nuclei coupled to layer one. Layer three contains all nuclei coupled to layer two which are not in layer one, and so on. This is illustrated in fig. \ref{pyramid}.

Each layer includes spins of only one species, layer one being all $A$, layer two all $B$, layer three all $A$ again and so on.  We envision taking only half the cube lattice, so that each layer is larger than the previous one. Layer $i$ has $i$ more spins than layer $i-1$.

For sake of analysis, suppose the crystal's Hamiltonian has nearest neighbor (Manhattan distance one)couplings only. In such a Hamiltonian, each A spin is coupled only to B spins, and \emph{vice versa}. The resonant frequency of each A (conversely B) spin is affected by the states it's B (conversely A) spin neighbors. Since the neighbors are all indistinguishable spins, only the total \emph{field} value is important, i.e. the number of neighbors pointing up, minus the number of neighbors pointing down.

Using the methods first suggested by Lloyd \cite{lloyd93}, also used in \cite{benjamin1} and others, it is possible to create gates that address species A spins that have a particular neighbor field value. For instance, it is possible to apply a $NOT$ gate (using the language of quantum computation) to all A spins that have neighbor field value $0$, \emph{and only} these spins.

We now show how to use the lattice structure to \emph{`amplify'} the spin we wish to measure  to a  detectable signal strength. We call the set of all species X spins, with neighbor field $k$, $X_k$.

The \emph{algorithm} is the following. Repeatedly apply a NOT gate to the following sets of spins: $B_{-2}$, $B_{-1}$, $B_{0}$, $A_{-2}$, $A_{-2}$, $A_{0}$. Each NOT gate consists of a (constant size) sequence of strongly modulated pulses.

Suppose that $\alpha = 1$, i.e. $\ket{\psi} = \ket{-1}$. Then absolutely nothing happens to the lattice (ignoring errors), and so it remains in the all $\ket{-1}$ state. To see this, note that all lattice points have at least three neighbors, all in the \ket{-1} state. Hence the field $k$ of every cell is at most $-3$. Since we are only doing flips on $k= 0, -1, -2$ this does not affect the lattice at all.

Suppose now that $\beta = 1$, i.e. the top cell of the pyramid is initialized to the state $\ket{+1}$.  Then, all $B$ neighbors of this vertex (those in layer 2) will have field value $-2$: $-3$ from three downward neighbors each in state $\ket{-1}$ and a $+1$ from the upward neighbor in state $\ket{+1}$. Therefore they will be flipped on the first $\pi$ pulse. No other nuclear spins will be affected. In the next $A$ stage, all $A$ cells in layer 3 will be flipped, and so on. In $n$ stages all cells in the first $n$ layers will be flipped to $\ket{+1}$. 

After $n$ stages there will be $N = \frac{1}{6}(n+1) n (n-1)$ nuclear spins pointing up. Hence, only $O(\sqrt[3]{N})$ stages are required to obtain a total field of $N$ spins.

By linearity, in the general case where  $\ket{\psi} = \alpha \ket{-1} + \beta \ket{+1}$, in the complete absence of decoherence the procedure would result in a cat state of $N$ nuclear spins in the state $\alpha \ket{-1}^{\tensor N} + \beta \ket{+1}^{\tensor N} $.

At the end of such procedure, one can measure the magnetic field of the cube lattice crystal, obtaining the desired measurement result.

Khitrin \emph{et. al.} used a similar approach to ours  \cite{Khitrin} in their scheme for polarizing spin chains. Although their method could potentially be used for spin measurement, the method presented here is cubically more efficient.

\begin{figure}
\includegraphics[scale=1]{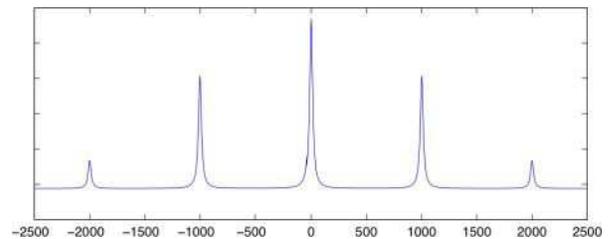}
\caption{\label{ideal} \emph{Ideal NMR Spectrum:} This is the ideal spectrum for a spin on the second layer. There are 5 distinct peaks, one for each of the five neighbor field values: $-4$, $-2$, $0$, $+2$, and $+4$. 
}
\end{figure}

In order for the scheme to work it is tantamount that the frequencies of the target spins be different from the ones we do not wish to affect. It would also be highly desirable to have the resonant frequencies of all target spins lumped together, as opposed to interspersed with the non-target frequencies. In the ideal setting, where the Hamiltonian has only first neighbor (strictly speaking Manhattan distance $1$) couplings, this is the case. In figure \ref{ideal} we see what an ideal spectrum for a second layer spin looks like. It has 5 distinct peaks, one for each of it's possible neighbor fields: $-4$, $-2$, $0$, $+2$, and $+4$.

We have so far assumed a nearest neighbor coupling Hamiltonian. A solid crystal, however, is governed by a direct dipole coupling Hamiltonian,
\[ \mathcal{H} = \sum_{i < j} d_{i,j} \left[ 2 \sigma_z^{(i)}\sigma_z^{(j)}  - k_{i,j}\frac{1}{2}\left( \sigma_+^{(i)}\sigma_-^{(j)} + \sigma_-^{(i)}\sigma_+^{(j)}\right)\right], \]
where $k_{i,j}$ equals one if $i,j$ are of the same species, and zero otherwise. And
\[   d_{i,j} = \frac{ g_{i,j} }{r_{i,j}^3} \frac{1}{2} \left( 3 \cos \Theta_{i,j} -1 \right),  \]
where $r_{i,j}$ is the distance between the two nuclei, $ \Theta_{i,j}$ is the angle between the vector connecting the two nuclei and the z-axis (determined by the magnetic field) and $ g_{i,j}$ is a simple constant that depends only on the nuclear types of $i$ and $j$.

\begin{figure}
\includegraphics[scale=1]{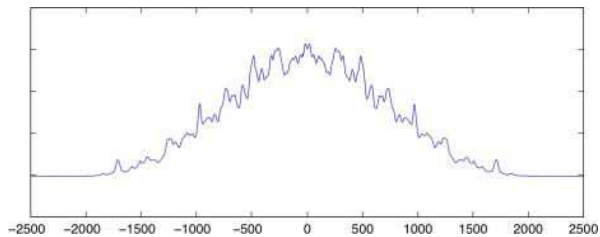}
\caption{\label{angle60} \emph{Rhombohedral lattice NMR Spectrum:} This is the spectrum for a spin on the second layer of a rhombohedral lattice with angles $\alpha = \beta = \gamma = \pi /3$.  The non-nearest neighbors, and in particular, the homonuclear couplings, swamp out the peaks. It is impossible to discern and address the wanted spins.
}
\end{figure}

Since the angle between the edges of the lattice and the z-field affect the coupling strength, it makes sense to orient the crystal such that the top of the pyramid points up in the z direction, making the angles symmetric. The problem is that, in a cubic lattice (with the Bravais angles $\alpha = \beta = \gamma = \pi /2$), all edges in the lattice have an angle to the z pole of exactly $\arctan \sqrt{2}$. This is known as the magic  angle. At this angle the coupling becomes zero.

It is clear that a cubic lattice is inappropriate if we are in a dipole coupling regime, and we wish to have the angles symmetric. A simple solution is to use a rhombohedral  crystal structure. We take one where the Bravais angles are $\alpha = \beta = \gamma = \pi /3$. In figure \ref{angle60} we show the spectrum of a second layer spin. Same as before, the pyramid peak points up in the z direction.  Although the nearest coupling are now there, they are washed away by the homonuclear couplings and non-nearest neighbor couplings. It is clear that we cannot address the spins we wish.

\begin{figure}
\includegraphics[scale=1]{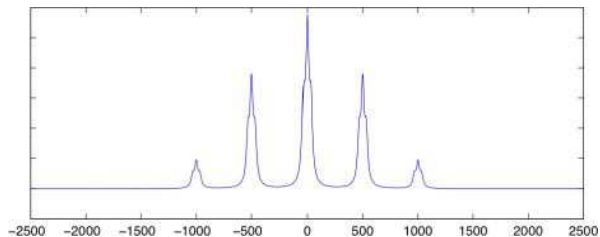}
\caption{\label{angle60s} \emph{Suppressing homonuclear coupling:} This is the spectrum for a spin on the second layer of a rhombohedral lattice with angles $\alpha = \beta = \gamma = \pi /3$, with the homonuclear couplings suppressed.  Note the similarity to the ideal case.
}
\end{figure}

However, it is possible to suppress all homonuclear couplings(c.f. \cite{cory90}). When we do so we get a Hamiltonian that looks remarkably close to the ideal nearest neighbor one. In figure \ref{angle60s} we show the spectrum of a second layer spin. Notice how similar it is to the ideal case. We can show similar results for different Bravais angles; in conclusion, almost any crystal structure will work properly when we suppress homonuclear interactions. 

In all cases presented, the crystals are hypothetical constructs, where the  couplings are set up so that the nearest neighbor couplings are roughly 1000 Hz.

It is clear that imperfect polarization will reduce the efficiency of our algorithm, yet it is a necessary consideration. Also, we have so far assumed only nearest neighbor interaction, which is not usually the case. A weak coupling can be observed between spins at greater distances, not taking these into account will result in imperfections in our gates. It is important to deal with these issues.

One important fact to notice is that our scheme is impervious to phase-flip errors in the lattice. Effectively, all pulses, and the final output, depend only on the diagonal terms of the density operator $\rho$ of the lattice. This is critical to state, not only because it deals with the algorithm's robustness to errors, but also because  it allows us to do an approximate, non-coherent, simulation of the evolution of our algorithm, as we show below.

The diagonal elements of $\rho$ must be protected, in the language of quantum information this amounts to protecting $\rho$ against bit-flip errors. A single  bit-flip error in the cube lattice by itself does not drastically change the  overall field, however, if this bit flips during the running of the algorithm it has the potential to influence all lattice points in a radius around it during the next step of the algorithm, by changing their neighbor fields during the pulse sequence, potentially creating a cascading of errors effect.

The first fact to notice is that errors inside the pyramid are less critical than errors on the faces, or worse, edges, of the pyramid. The reason  is that inside the pyramid all lattice points have exactly 6 neighbors. If one flips erroneously its neighbors still have 5 neighbors each in the correct state.

Take an interior lattice point that is in the state $\ket{-1}$ and should be flipped to $\ket{+1}$ in the next pulse sequence. Since it is supposed to flip it should have three positive neighbors and three negative
ones. Suppose that an upward neighbor has an error, then it will have two positive and four negative. This is, however, not a problem. The lattice point in question will have a field value $k = -2$, instead of $k = 0$, but will still be flipped correctly.

Contrast this to the case where a bit flip occurs along the edge of the pyramid. In this case the field of the lower neighbors to a lattice point with error will go from having a field of $-2$ to a field of $-4$ and will no longer be correctly flipped at all. The error will then propagate downwards.

One way to reduce errors is to extend the flipping operation to spins with neighbor fields of $+1$. This will correct some edge-related errors. One other way to drastically reduce error propagation is to force errors into the inside of the pyramid.

\begin{figure}
\includegraphics[scale=1]{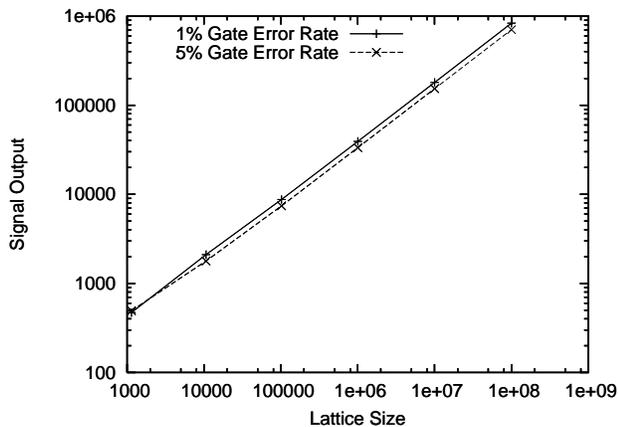}
\caption{\label{sizeplot} \emph{Output Signal:} This graph shows the signal strength at the end of the algorithm, as a function of the lattice size. The size is given in number of nuclei, and the signal strength is in number of correctly polarized nuclei. These numbers are for an initial crystal polarization of $90\% $ and gate error rates one and five percent. }
\end{figure}

Previously we concluded that we need to suppress the homonuclear coupling during our pulse sequences. In this Hamiltonian, the term $\sigma_+^{(i)}\sigma_-^{(j)} + \sigma_-^{(i)}\sigma_+^{(j)}$, often referred to as the \emph{`flip-flop'} term, has the effect of swapping two anti-aligned neighboring spins. Take a horizontal slice of the pyramid and suppose there is a single error among these lattice points. The effect of the flip-flop term on neighboring spins is well known and is called \emph{spin diffusion},  \cite{waugh}.  A first order approximation to this dynamic is a quantum walk on the lattice. In this model, an erroneous bit not only becomes diffused over several spins. Also, it  has a much lower probability (averaged over time) of being measured in non-interior points. This leads us to believe that letting the system evolve under  the homonuclear coupling for some time before running the algorithm will have the effect of reducing errors in the edges, and increase overall robustness to errors.

In order to test the reliability of our algorithm in the large scale it would be infeasible to have a coherent simulation. However, as stated above, the non-coherent nature of our procedure allows us to take a different approach. One simulation we set up is a non-coherent  Monte Carlo style simulation. We set up a three dimensional byte array, each byte representing a nuclear spin. Each byte is set to an initial value of $-1$. In order to simulate an initial polarization of $1- \epsilon_0$ we then flip the sign of each byte with probability $\epsilon_0$. Each pulse is simulated in a similar fashion. A gate error $\epsilon_1$ is simulated by failing to flip bytes with probability $\epsilon_1$. This simulation is repeated a number of times in order to get a meaningful statistical average. This technique allows to simulate very large systems, of size up to $10^8$ nuclear spins. 

We calculated the signal strength at the end  of our algorithm under different conditions of initial crystal polarization, and pulse error rate. In figure \ref{sizeplot} we plot the signal strength as a function of lattice size, when the initial polarization is $90\%$ and pulse error rates are $1\%$ and $5\%$. As can be seen from the graph, in order to get a signal strength in the order of $10^6$ polarized spins it is necessary to use a crystal lattice with about  $10^8$ nuclei. Running our algorithm on such a lattice requires about $800$ gates.

In closing, moving to a three dimensional measurement system, as opposed to a $1D$ chain as in \cite{paola}, gives distinct advantages. The clearest benefit is a cubic speedup in the procedure. This speedup was expected. Also the algorithm gains substantial robustness to errors. This is an even greater benefit, given the current fidelity of NMR operations.   Our numerical simulations tell us that the scheme  is highly efficient and robust. However, the nature of our simulations do not allow us to take into account large coherent superpositions or errors throughout the pyramid lattice. Large scale coherent simulations are infeasible. Ultimately we will need an actual physical implementation in order to best understand the  capabilities, and limits, of our proposed scheme.

This paper was written, in part, with funding from ARDA, ARO, CFI, CIAR, DARPA, LPS, MITACS, NSERC, and ORDCF.


\end{document}